\documentclass[12pt]{article}

\usepackage{graphicx}  

\title{Astrophysical constraints on hypothetical 
variability of fundamental constants}

\author{Sergei A. Levshakov}
\date{ }

\begin{document}

\maketitle              

\begin{center}
Department of Theoretical Astrophysics, 
Ioffe Physico-Technical Institute, \protect\newline 
Politekhnicheskaya Str. 26, 194021 St. Petersburg, Russia
\end{center}

\begin{abstract}
Many-multiplet (MM) method \cite{d99}
\index{Many-multiplet~method}
applied to three inhomogeneous samples of Keck/HIRES quasar absorption spectra
gives a shift in the value of the fine-structure constant of 
\index{Variation~of~fundamental~constants!alpha@$\alpha$}
$\Delta\alpha/\alpha = (-5.4\pm1.2)\times10^{-6}$ in the redshift range
$0.2 < z < 3.7$~\cite{m03}.
The 1~$\sigma$ error claimed in~\cite{m03} is, however, much too small
and cannot be maintained by current observations of quasars.
We present a modified MM method
to set an upper limit on $\Delta\alpha/\alpha$
from a homogeneous sample of Fe\,{\sc ii} lines identified in
the up-to-date best quality VLT/UVES spectrum of HE~0515--4414.
Our result is
$\Delta\alpha/\alpha = (1.1\pm1.1)\times10^{-5}$ at $z = 1.149$.
Theoretical models of the fundamental physical interactions predict
that the proton-to-electron
\index{Variation~of~fundamental~constants!memp@$m_e/m_p$}
mass ratio ($\mu = m_{\rm p}/m_{\rm e}$)
may relate to the shift in $\Delta\alpha/\alpha$ as
$\Delta\mu/\mu = {\cal R}\,\Delta\alpha/\alpha$.
We use VLT/UVES high-resolution observations of molecular hydrogen H$_2$
\index{molecular~hydrogen}
ultraviolet absorption lines at $z = 3.025$ toward Q~0347--3818
to bound the value of ${\cal R}$.
The obtained constraints on $\Delta$~ln~$\mu = (2.1\pm3.6)\times10^{-5}$ and
on  $\Delta$~ln~$\alpha = (1.1\pm1.1)\times10^{-5}$ rule out vary
large values of $|{\cal R}| > 6$.
Future observations with a new
High Accuracy Radial velocity Planet Searcher (HARPS) spectrograph
may provide a crucial test for the
$\Delta\alpha/\alpha$ measurements at a level of $10^{-6}$.
\end{abstract}

\section{Introduction}

The purpose of this paper is to outline some problems related to
the measurements of the dimensionless fundamental constants 
from quasar (QSO) 
absorption-line spectra:
the fine-structure constant $\alpha = e^2/\hbar c$ 
and the proton-to-electron mass ratio $\mu = m_{\rm p}/m_{\rm e}$, 
which characterize the strength of electromagnetic and strong interactions,
respectively.

The current interest in this field
stems from the prospect of using the high-resolution spectra
of extragalactic objects to measure with high accuracy
the transition energies of
a certain ionic species and molecules 
at different cosmological 
redshifts $z$\footnote{
$z = (\lambda - \lambda_0)/\lambda_0$, where $\lambda$ is the observed
wavelength of a given ionic transition and $\lambda_0$ is the
corresponding rest frame (laboratory) value. Modern observational data
are ranging from $z = 0$ (the Milky Way) up to
$z \sim 6$  (distant QSOs).}. 
If physical constants varied in the past, then
the comparison of the transition energies measured from QSO spectra
with their laboratory values provides direct astrophysical
constraints on the ratios $\alpha_z/\alpha$ and $\mu_z/\mu$ over a cosmological
time-scale ($\Delta t \sim 10^{10}$ yr).
We define $\Delta\alpha/\alpha = (\alpha_z - \alpha)/\alpha$ and
$\Delta\mu/\mu = (\mu_z - \mu)/\mu$, where index $z$ marks the values
of the fine-structure constant and the proton-to-electron mass ratio 
at redshift $z$, whereas their laboratory values
at $z = 0$ are denoted by $\alpha$ and $\mu$, respectively\footnote{In
the literature $\mu$ is sometimes referred to its inverse value, 
$\mu^\ast = m_{\rm e}/m_{\rm p}$. In this case $\Delta\mu^\ast/\mu^\ast = -
\Delta\mu/\mu$.}.
At present, these constants are measured with the relative accuracies 
$\delta_\alpha = 4\times10^{-9}$ and $\delta_\mu = 2\times10^{-9}$
and their values are equal to 
$\alpha = 1/137.03599958$ and $\mu = 1836.1526670$~\cite{mt00}.

The variability of the fundamental constants was firstly suggested by
Milne~\cite{m37} and Dirac~\cite{d37}
who assumed that the Newtonian gravitational constant $G$ 
may change in cosmic time.
Different functional forms for the time variations of $G$ and 
other constants were later
considered in a series of publications (see~\cite{u03}, for a review).
The variability of 
$\alpha$ caused by a coupling between electromagnetism 
and gravity is considered
within the framework of different modifications of the
Kaluza-Klein theory (e.g., \cite{cd80}-\cite{mb03}). 
The number of theoretical publications on this topic 
is considerably increased in the last two years.

For astronomical observations, it is more convenient to deal with
dimensionless constants since they are independent on the choice of
physical units. That is why most efforts to 
reveal hypothetical variability of constants are connected with measurements
of $\alpha$ 
(see~\cite{m03,b03} and references cited therein).

First observational study (1967) of the time dependence of $\alpha$
from QSO absorption spectra 
was based on the analysis of alkali doublets (AD method)
and yielded $\Delta \alpha/\alpha = -0.02\pm0.05$ at $z = 1.95$~\cite{bss67}.
Since then, there have been made many other AD studies 
of $\Delta \alpha/\alpha$ with increasing accuracy in
accord with increasing quality of observations: 
$\Delta \alpha/\alpha = (0\pm1.5)\times10^{-3}$ at $z = 3.2$~\cite{vp94},
$\Delta \alpha/\alpha = (-4.5\pm4.3\pm1.4)\times10^{-5}$ at $z = 2-4$~\cite{var00},
$\Delta \alpha/\alpha = (-0.5\pm1.3)\times10^{-5}$ at $z = 2-3$~\cite{m01}.
However, the use of absorption lines from different multiplets in different ions,
the so-called many-multiplet (MM) method,
may in principle increase the accuracy of
these $\Delta\alpha/\alpha$ estimations \cite{d99,m03}. 

Applying the MM method to
three samples of Keck/HIRES QSO data,  Murphy et al.~\cite{m03} obtained
a statistically significant negative value of
$\Delta \alpha/\alpha = (-5.4\pm1.2)\times10^{-6}$ 
in the redshift interval $0.2 < z <  3.7$.
In spite of all advantages of the MM method,
the result is nevertheless surprising 
since it implicitly assumes extremely stable conditions during the observing
nights and unrealistically homogeneous physical parameters in the absorption
clouds. 
We consider computational
caveats of the AD and MM methods in Section~2.

It should be emphasized that recent improvements in meteoritic abundance determinations
lead to the bound $\Delta\alpha/\alpha < 3\times10^{-7}$ at $ z = 0.45$~\cite{No5}
which contradicts the MM measurements~\cite{m03} at the same redshift.

If $\alpha$ is supposed to be time dependent, the other gauge coupling
constants should also depend on time as predicted by
theoretical models of the fundamental physical interactions
(e.g., \cite{No1}-\cite{No8}).
However, the relation between the variations of the low energy gauge couplings
is highly model dependent.
For instance, within the framework of a unified theory
[e.g., supersymmetric SU(5)] the time variation of
the QCD scale parameter $\Lambda$ leads to the change of the proton mass
$m_{\rm p}$ which is proportional to the shift
$\Delta\alpha/\alpha$~\cite{No1,No2,No3}:
\begin{equation}
\frac{\Delta m_{\rm p}}{m_{\rm p}} = \frac{\Delta \Lambda}{\Lambda} =
{\cal R}\,\frac{\Delta \alpha}{\alpha}\; ,
\label{10E1}
\end{equation}
where the factor ${\cal R}$
is known with large theoretical uncertainties~\cite{No6,No7,No8}.

Theoretically predicted numerical value of ${\cal R}$
can be tested by measuring
the electronic-vibrational-rotational
lines of the H$_2$ molecule observed in QSO spectra at high-$z$
as described in~\cite{vl93}. We discuss this test in Section~3.
Our conclusions and remarks for 
future prospects are summarized in Section~4.

\section{Methods to constrain $\Delta\alpha/\alpha$ from QSO absorption spectra}

\subsection{The alkali-doublet (AD) method}

The value of the fine-structure constant $\alpha_z$ at redshift $z$
can be directly estimated from the relative wavelength separation between
the fine-splitting lines of an alkali doublet (AD method)~\cite{bs67}: 
\begin{equation}
\frac{\alpha_z}{\alpha} =
\left( \frac{\Delta \lambda_z}{\langle \lambda \rangle_z}\left/
\frac{\Delta \lambda}{\langle \lambda \rangle}\right. \right)^{1/2} \; ,
\label{10E2}
\end{equation}
where $\Delta \lambda_z$ and $\langle \lambda \rangle_z$ are, respectively,
the fine-structure separation and the weighted mean wavelength for a given
doublet from a QSO absorption system at redshift $z$, while 
$\Delta \lambda$ and $\langle \lambda \rangle$ denote the same quantities 
from laboratory data.

\noindent
The uncertainty $\sigma_{\Delta\alpha/\alpha}$ in the individual measurement
can be estimated from~\cite{lev94}:
\begin{equation}
\sigma_{\Delta\alpha/\alpha} \simeq
\frac{1}{\sqrt{2}}\left(\delta^2_{\lambda_z} + \delta^2_\lambda\right)^{1/2}\: ,
\label{10E3}
\end{equation}
where $\delta_{\lambda_z} = \sigma_{\lambda_z}/\Delta\lambda_z$,
$\delta_\lambda = \sigma_{\lambda}/\Delta\lambda$, and $\sigma_{\lambda_z}$, 
$\sigma_{\lambda}$ are, respectively, the errors of the wavelengths 
in the absorption system and laboratory.

Equation~(\ref{10E2}) shows that the AD method uses only the wavelength
ratios and thus it does not require the measurements of absolute wavelengths.
This fact excludes large systematic errors caused by the wavelength
calibration in a wide spectral range. 
Therefore the accuracy of the AD measurements is
mainly restricted by statistical fluctuations in the recorded counts at different
wavelengths within the line profile. 

The measurement of the line profile parameters and their errors are
thoroughly described in~\cite{bohl}.
The midpoint $\lambda_c$ of an interval over $M$ pixels covering
the line profile can be calculated from the equations
\begin{equation}
\lambda_c = \frac{a}{W_\lambda}\,\sum^M_{i=1} \lambda_i(1 - I_i/C_i)\: ,
\label{10E4}
\end{equation}
and
\begin{equation}
W_\lambda = a\,\sum^M_{i=1} (1 - I_i/C_i)\: ,
\label{10E5}
\end{equation}
where $W_\lambda$ is the equivalent width, $a$ is the wavelength interval between
pixels, $I_i$ is the observed count rate at pixel $i$, $C_i$ is the continuum count
rate at pixel $i$, and $\lambda_i$ is the wavelength at pixel $i$.

The  errors in $\lambda_c$ and $W_\lambda$
caused by counting statistic and the uncertainty
of the overall height of the continuum level are given by~\cite{bohl}:
\begin{equation}
\sigma_{\lambda_c} = \frac{a}{W_\lambda}
\left[
\left\langle \frac{\sigma_I}{C} \right\rangle^2 \left( 
A - 2\lambda_c B + M\lambda^2_c\right) +  
\left\langle \frac{\sigma_C}{C} \right\rangle^2\left(
B - M\lambda_c\right)^2 \right]^{1/2},
\label{10E6}
\end{equation}
and
\begin{equation}
\sigma_{W_\lambda} = a\,\left[
\left\langle \sigma_I/C \right\rangle^2\,M +
\left\langle\sigma_C/C \right\rangle^2 \left(
M - W_\lambda/a \right)^2 \right]^{1/2},
\label{10E6a}
\end{equation}
where $A = \sum_i\lambda^2_i$, $B = \sum_i\lambda_i$, $\langle \sigma_I/C \rangle$
is the inverse mean signal-to-noise 
$(S/N)^{-1}$ ratio over the line measurement interval, and
$\langle \sigma_C/C \rangle$ is the mean accuracy of the continuum level.

\noindent
If the rms error of the wavelength scale calibration is $\sigma_{\rm sys}$,
then the total error is given by the sum
\begin{equation}
\sigma^2_{\rm tot} = \sigma^2_{\lambda_c} + \sigma^2_{\rm sys}.
\label{10E7}
\end{equation}
The error of the wavelength scale calibration is related to the radial velocity
accuracy as
\begin{equation}
\sigma_v = c\,\sigma_{\rm sys}/\lambda\; ,
\label{10E7a}
\end{equation}
where $c$ is the speed of light. 

For example, $\sigma_{\rm sys} \simeq 0.002$ \AA\, for VLT/UVES data~\cite{lddm}
corresponds at 5000~\AA\, to $\sigma_{\rm sys} = 0.12$ km~s$^{-1}$.
If $\sigma_{\lambda_c} = 0.002$ \AA, then for the Mg\,{\sc ii} $\lambda\lambda
2796, 2803$ \AA\, doublet (\ref{10E3}) gives
$\sigma_{\Delta\alpha/\alpha} = 1.3\times10^{-4}$ at $z = 2$, which is 
a typical accuracy for the individual AD estimations from isolated doublets
with $W_\lambda \sim 100$ m\AA.
Equation~(\ref{10E6}) shows 
that $\sigma_{\lambda_c}$ decreases linearly with $W_\lambda$,
meaning that, not surprisingly, the error for weak absorption lines 
($W_\lambda \ll 1$ \AA) may be very large. For example, the rest-frame
equivalent widths for Fe\,{\sc ii} $\lambda1608$~\AA\, and $\lambda1611$~\AA\,
at $z = 3.39$ toward Q0000--2620 measured from the Keck/HIRES data are
$W_{1608} \simeq 0.178$ \AA\, and $W_{1611} \simeq 0.003$ \AA~\cite{lu96}, i.e.
the measured errors $\sigma_{\lambda_c}$ for these lines should differ by
$\sim 50$ times. It is clear that 
to achieve the highest accuracy in the $\Delta\alpha/\alpha$ estimations
by means of the the AD method only well separated and unsaturated doublets
with $W_\lambda \sim 100$ m\AA\, should be selected for the analysis.

\subsection{The many-multiplet (MM) method}

The many-multiplet (MM) method is a generalization of the AD 
method (see~\cite{m03} and references therein).
The method uses many transitions from different multiplets and
different ions observed in a QSO absorption system. 
The line profiles selected for the MM analysis are represented
by a sum of the same number of components (Voigt profiles) each of which is
defined by three individual parameters (the column density, the Doppler width
or $b$-parameter, and the redshift $z_i$) and a common value of
$\Delta\alpha/\alpha$.   
The rest wavenumbers in the MM method can be altered according to~\cite{m03}
\begin{equation}
\omega_z = \omega_0 + q\,x_z\; ,
\label{10E9}
\end{equation}
where
\begin{equation}
x_z = (\alpha_z/\alpha)^2 - 1\; ,
\label{10E10}
\end{equation}
and $\omega_z$, $\omega_0$ are the rest-frame wavenumbers of a transition in a QSO
system with redshift $z$ and measured in the laboratory, respectively.
The sensitivity coefficients $q$ (determining the sensitivity of
$\omega_0$ to the variation of $\alpha$) are listed
in Table~2 in~\cite{m03}.

All parameters are estimated through the minimization of $\chi^2$
simultaneously for the whole set of lines.
A significant improvement in the accuracy of the $\Delta\alpha/\alpha$
estimations by the MM method compared to the AD method is due to a 
wide range of the $q$ values (see Section~2.3 below).
Nevertheless, the MM method has some immanent shortcomings which could
affect the claimed accuracy 
$\sigma_{\langle \Delta\alpha/\alpha \rangle} = 1.2\times10^{-6}$.

First of all, the weakest point of the MM method is the 
multiple velocity component Voigt profile deconvolution.
From mathematical point of view,
any deconvolution itself is a typical \emph{ill-posed} problem~\cite{lta99}.
When several smoothing operators (like convolution
with the spectrograph point-spread function, summing, and integrating) 
are involved in the minimization procedure, 
the ill-posing may cause quite ambiguous results.
In case of complex line profiles when many components are required
to describe the whole profile, 
the fitting parameters cannot be estimated with high accuracy because of
strong \emph{inter-parameter correlations}. 
For example, the value of $\Delta\alpha/\alpha$ 
can be artificially constrained by the increase of the component
number or by variations of $b$-values. 
Thus, to enhance the reliability of the results only simple line
profiles (requiring minimum number of components) should be taken into
MM analysis.

From physical point of view, the Voigt profile deconvolution assumes
that the complex profile is caused by separate clouds lined along the
line of sight with different radial velocities, each having its own
constant gas density, kinetic temperature 
and, hence, its own ionization state.
However, observations of QSOs have shown that with increasing spectral
resolution more and more subcomponents appear in the line profile. This
complexity indicates that, in general, metal lines arise in continuous turbulent media
with fluctuating velocity and density fields. 

We note in passing that the highest spectral resolution 
achieved in observations of QSOs at modern giant telescopes is
$\sim 5$ km~s$^{-1}$ ($FWHM$), whereas 
the expected values for minimum (thermal) widths of metal lines
are less or about 1~km~s$^{-1}$
for typical kinetic temperatures 100~K $< T_{\rm kin} < 10^4$~K. 
This means that we are still 
not able to measure directly the true intensity in the line
profile but observe only the apparent intensity which is a convolution of the true
spectrum and the spectrograph point-spread function.

Thus, to measure the
transition energies, the distribution function of the radial velocities
within the intervening cloud must be known for the AD method, whereas
the MM method requires in addition the knowledge of the density distribution,
for it deals with ions of different elements which may not arise 
co-spatially~\cite{lak00}. Since the velocity and density distributions are
determined from the same line profiles which are used in the
measurements of $\Delta\alpha/\alpha$, it is not possible to discriminate
among kinematic effects and possible changes in $\alpha$ on the basis of only 
one absorption system. 
But if we have many absorption systems,
statistical approach may probably solve this problem if 
shifts in the measured $\Delta\alpha/\alpha$ values are of random nature
and, thus,  can be averaged out in a large ensemble of absorption
systems where any deviations in velocity distributions of different ions
are equally probable.

Shifts in $\Delta\alpha/\alpha$ may also arise if the main
assumption of the MM method that different ionic species trace exactly
each other in their space distribution is not adequate.
Dissimilarities in the line profiles of different ions may be caused by
density fluctuations leading to inhomogeneous photoionization in the 
absorption cloud. 

The QSO absorption clouds is a very inhomogeneous population.
Their linear sizes, $L$, can be estimated
from the observed neutral hydrogen column densities, $N$(H\,{\sc i})
$\sim 10^{20} - 10^{21}$ cm$^{-2}$, and the estimated gas number densities,
$n_{\rm H} \sim 10^{-2}-1$ cm$^{-3}$, giving
$L = N$(H\,{\sc i})$/n_{\rm H} \sim$
10s pc -- 10s kpc. The intervening clouds may also contain small dense filaments
with much higher densities ($n_{\rm H} \sim 10^2$ cm$^{-3}$) and
the line-of-sight sizes less than 
1~pc~\cite{rbql}.

It is usually assumed that the metal absorbers in QSO spectra are in
photoionization equilibrium with the ambient metagalactic UV background.
To describe the thermal and the ionization state of the gas in the diffuse
clouds it is convenient to introduce the dimensionless `ionization parameter'
$U = n_{\rm ph}/n_{\rm H}$, equal to the ratio of the number of photons in
1 cm$^3$ with energies above one Rydberg to the gas number density. 

For illustrative purpose, we will assume that the UV background is
not time or space dependent and that its shape 
and intensity is given by the metagalactic
spectrum at $z = 3$ presented in~\cite{hm} ($n_{\rm ph} \simeq 2\times10^{-5}$
cm$^{-3}$). Then fractional ionizations $\Upsilon_{a,i} = n_{a,i}/n_a$ (with
$n_a$ being the local number density of element `$a$' and $n_{a,i}$ the
density of ions in the $i$th ionization state)
of different elements can be calculated as functions of $U$ using the
photoionization code CLOUDY~\cite{fer97}.

\begin{figure}[t]
\begin{center}
\includegraphics[width=.85\textwidth]{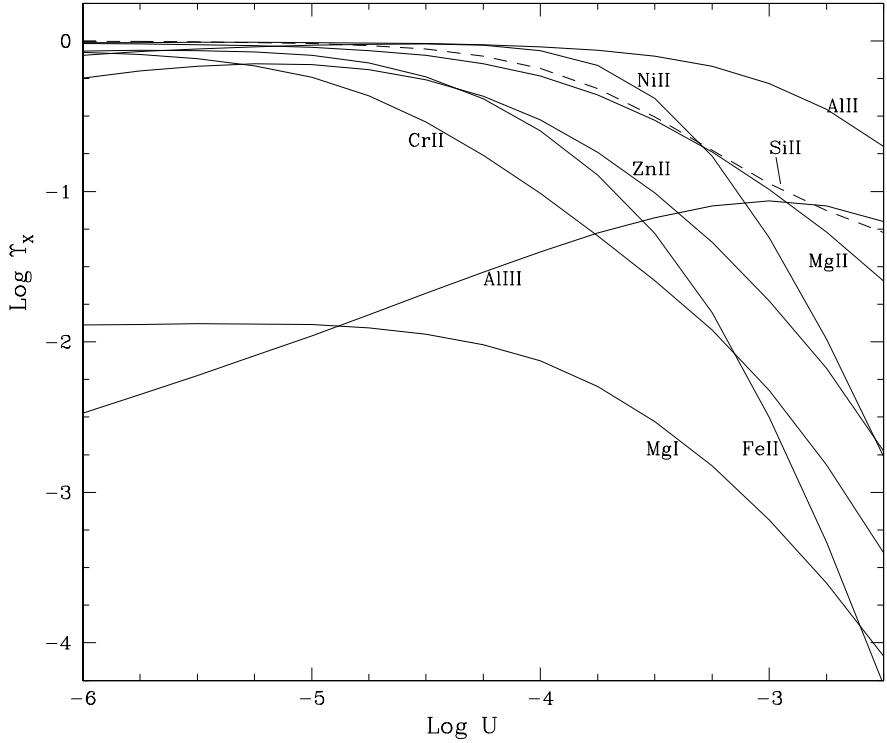}
\end{center}
\caption[]{Fractional ionizations $\log \Upsilon_X$ vs. the ionization
parameter $\log U$ in case of the Haardt-Madau~\cite{hm} metagalactic
UV background at $z = 3$ and metallicity $Z = 0.1Z_\odot$.
Similarity of the line profiles of a pair of
ions `$a$' and `$b$' requires the fractional ionization ratio
$\Upsilon_a/\Upsilon_b$ to be constant for different $U$ }
\label{10F1}
\end{figure}

Figure~1 shows these results for a metallicity $Z = 0.1Z_\odot$
(typical for the QSO metal absorbers) and for the elements involved in the
MM analysis in~\cite{m03}. Similar profiles for different ions can only
be observed when the ratios $\Upsilon_{a,i}/\Upsilon_{b,j}$ are constant,
i.e., the curves in Fig.~1 are parallel. This condition is realized for
Mg\,{\sc i},  Mg\,{\sc ii},  Al\,{\sc ii}, Si\,{\sc ii}, Fe\,{\sc ii},
Ni\,{\sc ii}, and Zn\,{\sc ii} in the range $\log U < -4.2$, i.e., for the
most dense volumes of the cloud. Such dense clumps (filaments) have,
however, a negligible filling factor since their sizes are small, and, thus,
the most of the diffuse cloud is filled up by a rarefied, low density
gas characterized by higher values of $\log U$. In the range
$\log U > -4$, the diversity of the $\Upsilon_X$ values is considerable and
one may expect to observe nonsimilar profiles of the above mentioned ions. 
Under these circumstances, in order to be more secure the MM method should
work with homogeneous samples of ions which have undoubtly 
the same volume distribution like, e.g., Fe\,{\sc ii} lines.

Note that
Al\,{\sc iii} is obviously 
not a good candidate for MM calculations since
the shape of the $\Upsilon_{{\rm Al}\,{\scriptscriptstyle\rm III}}$
curve clearly differs from the others.
The fact that no additional scatter in $\Delta\alpha/\alpha$ for systems
containing Al\,{\sc iii} was found in~\cite{m03} may imply that 
the available spectral resolution is not high
enough to obtain the true profiles of the metal lines.

To conclude this section we comment on the accuracy 
$\sigma_{\Delta\alpha/\alpha}$ which can be in principle
achieved from the up-to-date QSO observations. 
Using the method of error propagation, one finds from (\ref{10E9})
the relative error of $\omega_z$:
\begin{equation}
\delta_{\omega_z} \equiv \sigma_{\omega_z}/\omega_z = 2\,|Q|\,
\sigma_{\Delta\alpha/\alpha}\; .
\label{10E11}
\end{equation}
Here $Q = q/\omega_0$ is the dimensionless sensitivity coefficient.

For the lines listed in Table~2 in~\cite{m03}, the $Q$ values are ranging
from $-0.027$ (Cr\,{\sc ii} $\lambda2066$ \AA) to 0.050
(Zn\,{\sc ii} $\lambda2026$ \AA), however, for the most frequently used
transitions in the MM calculations (Table~1 in~\cite{m03}), 
this range is narrower:
$Q$(Mg\,{\sc ii} $\lambda2803$ \AA) = 0.0034, and 
$Q$(Fe\,{\sc ii} $\lambda2600$ \AA) = 0.035. 
These numbers show that the error of the sample mean
$\sigma_{\langle \Delta\alpha/\alpha \rangle} = 1.2\times10^{-6}$
found in~\cite{m03} requires the relative error
$\langle \delta_{\omega_z} \rangle \simeq 8.6\times10^{-9}-8.4\times10^{-8}$,
meaning that  the sample mean wavelength accuracy should be 
$\langle \sigma_\lambda \rangle \simeq 2\times10^{-5}-2\times10^{-4}$ \AA\,
at 2000 \AA\, ($\sigma_\lambda = \lambda\,\delta_{\omega_z}$), which is
equal to
the radial velocity accuracy of $\Delta v \simeq 2.5-25$ m~s$^{-1}$
($\Delta v = c\,\delta_{\omega_z}$). 
Even if the individual measurement is an order of magnitude less accurate
as compared to the sample mean, the uncertainty of $\Delta v \simeq 25-250$
m~s$^{-1}$ seems to be hardly realized for all systems collected in~\cite{m03}.
Namely, the thermal shifts of order 300 m~s$^{-1}$ are expected if
temperature changes on 0.5~K between the science and calibration spectra 
exposures~\cite{lddm}.

\begin{figure}[t]
\begin{center}
\includegraphics[width=1.0\textwidth]{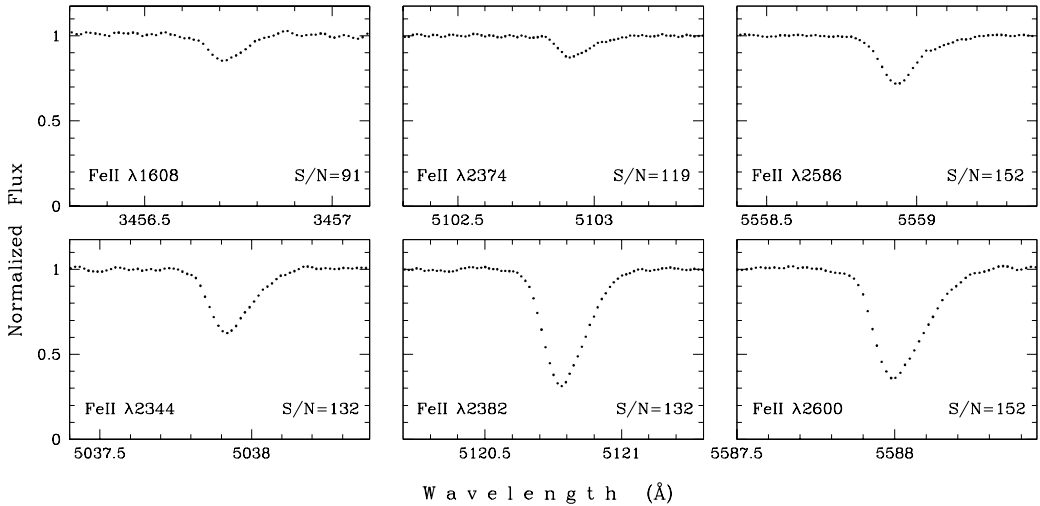}
\end{center}
\caption[]{Fe\,{\sc ii} lines associated with the $z = 1.149$ absorption
system toward the quasar HE~0515--4414 (normalized intensities are
shown by dots). The mean signal-to-noise (S/N) ratios per resolution
element are indicated
}
\label{10F2}
\end{figure}

\subsection{The regression MM method}

The standard MM method based on the multiple
velocity component Voigt profile fit to the complex spectra when 
$\Delta\alpha/\alpha$ is treated as an additional free parameter~\cite{m03}
can be easily modified to be free from the uncertainties mentioned in the
previous section. 
By its very nature, the MM method is similar to that  developed in \cite{vl93} for
the $\Delta\mu/\mu$ calculations. So, by analogy,
we can re-write (\ref{10E9})
in the linear approximation ($|\Delta\alpha/\alpha| \ll 1$)
in the form 
\begin{equation}
z_i = z_0 + \kappa_\alpha Q_i\; ,
\label{10E13}
\end{equation}
where the slope parameter $\kappa_\alpha$ is given by
\begin{equation}
\kappa_\alpha = 2\,(1 + z_0)\,\frac{\Delta\alpha^\ast}{\alpha}\; .
\label{10E14}
\end{equation}
Here $\Delta\alpha^\ast/\alpha = - \Delta\alpha/\alpha$.

\begin{table}[t]
\caption{Fe\,{\sc ii} lines at $z = 1.149$ toward the quasar HE~0515--4414
and sensitivity coefficients Q}
\begin{center}
\renewcommand{\arraystretch}{1.4}
\setlength\tabcolsep{5pt}
\begin{tabular}{lccccr}
\hline\noalign{\smallskip}
\multicolumn{1}{c}{$\lambda^a_0$, \AA} & $f$ & $\lambda_c,$ \AA & $W_\lambda$, m\AA 
& $z$ &\multicolumn{1}{c}{$Q$} \\
\noalign{\smallskip}
\hline
\noalign{\smallskip}
1608.45085(8)&0.0580$^b$&3456.7176(46)&7.8(1.8)&1.1490975(28)&$-0.0193(48)$ \\
2344.2130(1)&0.1140$^b$&5037.9398(23)&27.5(1.6)&1.1490964(10)&0.0294(35) \\
2374.4603(1)&0.0313$^b$&5102.9319(79)&8.4(1.5)&1.1490913(33)&0.0389(35) \\
2382.7642(1)&0.3006$^c$&5120.8015(22)&61.0(1.9)&1.1491012(9)&0.0357(36) \\
2586.6496(1)&0.0684$^d$&5558.9439(29)&20.1(1.4)&1.1490904(11)&0.0393(39) \\ 
2600.1725(1)&0.2239$^c$&5588.0269(21)&59.1(2.0)&1.1490985(8)&0.0353(39) \\
\hline
\multicolumn{6}{l}{$^a$Vacuum rest wavelengths are taken from~\cite{m03}.}\\
\multicolumn{6}{l}{REFERENCES: $^b$Welty et al.~\cite{w99}; $^c$Morton~\cite{mor91};
$^d$Cardelli \& Savage~\cite{cs95}.}\\
\end{tabular}
\end{center}
\label{10T1}
\end{table}

If $\Delta\alpha/\alpha$ is not zero, there should be a correlation between
$z_i$ ($= \lambda_{c,i}/\lambda_{0,i} - 1$) and $Q_i$, and the regression
parameters $z_0$ and $\kappa_\alpha$ can be estimated from the measured centers
of the isolated metal lines.

For instance, Fe\,{\sc ii} lines listed in Table~2 in~\cite{m03}, for which
the $Q$ values are ranging in the broad interval between $-0.019$ and 0.035, 
may be suitable for such calculations. 
Equation~(\ref{10E13}) shows that
the broader the interval of $Q$-values, the higher the accuracy of the slope
parameter $\kappa_\alpha$ (see, e.g., \cite{nr} where the errors of the
linear regression are calculated).

It is also an advantage that Fe\,{\sc ii} lines arise mainly in low ionized regions, 
where their thermal width is the 
smallest among other elements (Mg, Al, Si)  proposed for the MM method
(we do not consider Cr\,{\sc ii}, Ni\,{\sc ii}, and
Zn\,{\sc ii} as appropriate for the MM analysis because they are rare in
QSO spectra and, when observed, are usually very week) since the narrow lines
allow to measure the transition energies with highest accuracy.
Below we consider an example of such Fe\,{\sc ii} sample.

\begin{figure}[hpt]
\begin{center}
\includegraphics[width=.97\textwidth]{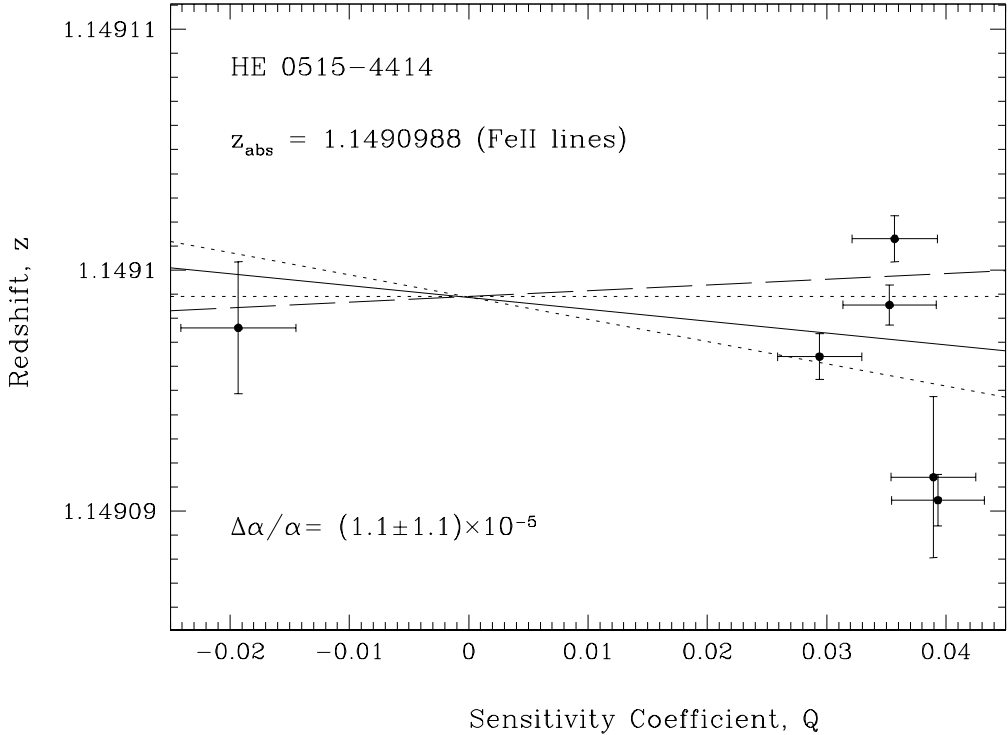}
\end{center}
\caption[]{Relation between the redshifts $z_i$ calculated for individual
Fe\,{\sc ii} lines listed in Table~1 and their sensitivity coefficients $Q_i$.
The solid line is the linear regression $z_i = z_0 + \kappa_\alpha\,Q_i$.
The dotted lines representing the 1~$\sigma$ deviations from the optimal slope
demonstrate that
$\Delta\alpha/\alpha = 0$ at the level $\sim 2\times10^{-5}$.
The dashed line is drawn for $\Delta\alpha/\alpha  = -5.4\times10^{-6}$
found in the MM analysis in~\cite{m03} 
}
\label{10F3}
\end{figure}

Figure~2 shows isolated and unsaturated Fe\,{\sc ii} lines from the metal
system at $z = 1.149$ toward the bright quasar HE 0515--4414. The spectra of
this quasar were obtained with the UV-Visual Echelle Spectrograph (UVES) installed
at the VLT/Kueyen telescope. The VLT/UVES data have a very high signal-to-noise
ratio ($S/N \simeq 90-150$ per resolution element) and a high resolution power
of $\lambda/\Delta\lambda \simeq 55\,000$ ($FWHM \simeq 5.5$ km~s$^{-1}$).
The absorption systems from this data were analyzed in \cite{rbql}
and \cite{qbr,larb}.

We used (\ref{10E4})-(\ref{10E7}) to estimate the rest-frame equivalent
widths $W_\lambda$, the line centers $\lambda_c$, and their errors which are
listed in Table~1. In this table, column~(2) presents the oscillator strengths
for absorption. The sensitivity coefficients $Q_i$ and their errors 
$\sigma_{Q_i}$ are calculated from the values $q_i$, $\sigma_{q_i}$, and
$\omega_{0,i}$ listed in Table~2 in~\cite{m03}.

Our result of the linear regression analysis is shown in Fig.~3 by the
solid line, while two dotted lines correspond to the 1~$\sigma$
deviations of the slope parameter $\kappa_\alpha$. We find
$z_0 = 1.1490988\pm0.0000016$ and $\Delta\alpha/\alpha = 
(1.1\pm1.1)\times10^{-5}$ (1~$\sigma$). The uncertainties of the $z_0$ and
$\Delta\alpha/\alpha$ values are calculated through statistical Monte Carlo
simulations assuming that the deviations in the $z_i$ and $Q_i$ values are
equally likely in the intervals
$[z_i-\sigma_{z_i}, z_i+\sigma_{z_i}]$ and
$[Q_i-\sigma_{Q_i}, Q_i+\sigma_{Q_i}]$. 
The result obtained shows that $\Delta\alpha/\alpha = 0$ within the
1~$\sigma$ interval. For comparison, the linear regression for
$\Delta\alpha/\alpha = -5.4\times10^{-6}$ based on the inhomogeneous sample
containing 127 absorption systems~\cite{m03} is shown by the dashed line in
Fig.~3. Although the sign of $\Delta\alpha/\alpha$ is opposite, the
value of $\Delta\alpha/\alpha = -5.4\times10^{-6}$ is consistent
with our data within the 2~$\sigma$ interval. The accuracy of the
regression MM analysis may be increased 
if a few homogeneous samples are combined. 
The regression analysis  is easily generalized
for this case (see, e.g.,~\cite{ivan}).

\section{Constraints on the proton-to-electron mass ratio}

The  proton-to-electron mass ratio $\mu$ can be estimated
from high redshift molecular hydrogen systems. 
With some modifications, such measurements
were performed for the $z = 2.811$ H$_2$-bearing cloud toward
the quasar PKS 0528--250 in~\cite{vl93} and \cite{pot98}
setting the limit of $|\Delta \mu/\mu|< 1.8\times10^{-4}$ (1~$\sigma$).
Later on higher resolution spectra of this QSO obtained with
the VLT/UVES~\cite{lckm03} 
revealed that the H$_2$ profiles at $z = 2.811$ are complex and
consist of at least two subcomponents with $\Delta v \simeq 10$ km~s$^{-1}$.
\begin{figure}[hpt]
\begin{center}
\includegraphics[width=.99\textwidth]{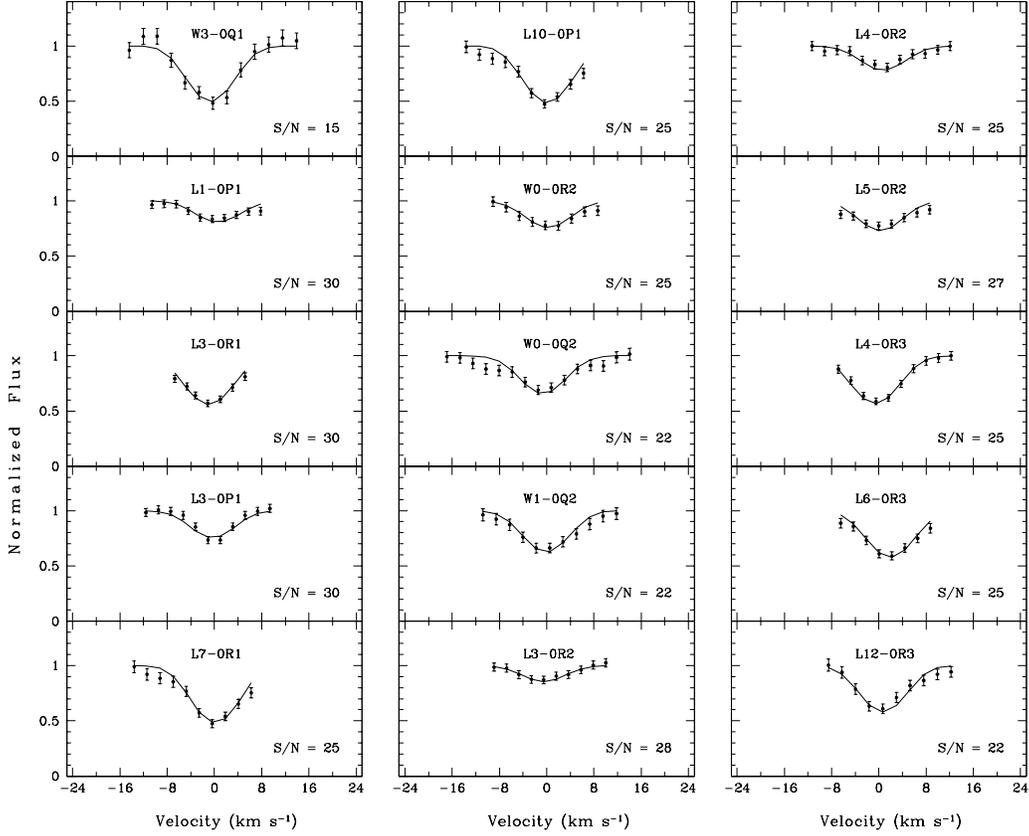}
\end{center}
\caption[]{H$_2$ absorption features identified at  $z = 3.025$ 
toward the quasar Q~0347--3819 (normalized intensities are shown 
by dots with $1\sigma$ error bars).
The zero radial velocity is fixed at $z_{{\rm H}_2} = 3.024895$. 
Smooth curves are the synthetic H$_2$ profiles found by the
least-squares procedure. 
The mean S/N values per resolution element
are indicated
}
\label{10F4}
\end{figure}
Lines arising from the low rotational levels with $J = 0$ and $J = 1$ 
show a red-side asymmetry -- contrary to those arising from $J = 2$ and 
$J = 3$ where a blue-side asymmetry is clearly seen.
This complexity makes impossible higher accuracy measurements of
$\Delta \mu/\mu$ in this system. Again, as in the case with 
the $\Delta\alpha/\alpha$
estimations, one needs a single component narrow H$_2$ absorption-line system
to restrict the observed wavelengths within the uncertainty interval of
a few m\AA.  

Such system has been found at $z = 3.025$ toward the quasar
Q0347--3819~\cite{lddm02}.
The absorption spectrum of this QSO was obtained with  
$FWHM \simeq 7$ km~s$^{-1}$ and 
$S/N \sim 15-30$ per resolution element at the VLT/UVES.
15 unblended H$_2$ lines  (selected from $\sim 80$ identified H$_2$
transitions) which
provide the most accurate line center measurements are shown in Fig.~4.

For a given H$_2$ line, the number of pixels involved in the analysis
corresponds to the number
of points in the line profile shown in Fig.~4.
To measure the line centers we used a method
that matches the observed profiles with the synthesized ones to estimate
a set of model parameters (for details, see~\cite{lddm}).
The results obtained are presented in Table~2.
The measured individual $z$ values and their standard deviations
are listed in column~(5). Using these redshifts
we can constrain possible changes of $\mu$.

The rest-frame wavelengths, $\lambda_{0,i}$, of different
electronic-vibrational-rotational transitions of H$_2$ depend
in a different way on the reduced mass of the molecule.
The sensitivity of $\lambda_{0,i}$ to variation of $\mu$
is given by the sensitivity coefficient ${\cal K}_i$
defined as~\cite{vl93}:
\begin{equation}
{\cal K}_i = \frac{\mu}{\lambda_{0,i}}\frac{d\lambda_{0,i}}{d\mu}\; .
\label{10E15}
\end{equation}

Following the procedure described in \cite{pot98}
we calculate the coefficients ${\cal K}_i$ listed in column~(6) of Table~2.
Their errors are obtained
from coefficients $Y_{mn}$ (presented in Table~1 in~\cite{pot98})
using the method of error propagation
($Y_{mn}$ values are considered to be accurate
to $k$ decimal places and their rounding errors
are set to $0.5\times10^{-k}$).

\begin{table}[hpt]
\caption{H$_2$ lines at z = 3.025 toward Q~0347--3819 and sensitivity
coefficients ${\cal K}$}
\begin{center}
\renewcommand{\arraystretch}{1.4}
\setlength\tabcolsep{5pt}
\begin{tabular}{llcclr}
\hline\noalign{\smallskip}
\multicolumn{1}{c}{$J$} & Line & $\lambda^a_0,$ \AA & $\lambda_c$, \AA 
& \multicolumn{1}{c}{ $z$} &\multicolumn{1}{c}{${\cal K}$} \\
\noalign{\smallskip}
\hline
\noalign{\smallskip}
$1\ldots$ & W3-0Q & 947.4218(5) & 3813.2653(41) & 3.024887(5) & 0.0217427(8)\\
& L1-0P & 1094.0522(51) & 4403.4575(77) & 3.02491(2) & $-0.0023282(1)$ \\
& L3-0R & 1063.4603(1)  & 4280.3051(30) & 3.024885(3) & 0.0112526(4)\\
& L3-0P & 1064.6056(5)  & 4284.9249(33) & 3.024894(4) & 0.0102682(4)\\
& L7-0R & 1013.4412(20) & 4078.9785(25) & 3.024898(9) & 0.03050(1)\\
& L10-0P&  982.8340(6) &  3955.8049(51) & 3.024896(6) & 0.04054(6)\\
\noalign{\smallskip}
$2\ldots$ & W0-0R & 1009.0233(7) & 4061.2194(61) & 3.024901(7) & $-0.0050567(7)$\\
& W0-0Q & 1010.9380(1) & 4068.9088(53) & 3.024885(6) & $-0.0068461(2)$\\
& W1-0Q & 987.9744(20) & 3976.4943(49) & 3.02490(1) & 0.0039207(2)\\
& L3-0R & 1064.9935(9) & 4286.4755(98) & 3.02488(1) & 0.0097740(5)\\
& L4-0R & 1051.4981(4) & 4232.1793(85) & 3.024904(8) & 0.015220(1)\\
& L5-0R & 1038.6855(32) &4180.6048(70) & 3.02490(1) & 0.020209(3)\\
\noalign{\smallskip}
$3\ldots$ & L4-0R & 1053.9770(11) &4242.1419(28) & 3.024890(5) & 0.012837(2)\\
  & L6-0R & 1028.9832(16) &4141.5758(31) & 3.024921(7) & 0.022332(7)\\
  & L12-0R& 967.6752(21) &3894.7974(39) & 3.02490(1) & 0.0440(2) \\ 
\hline
\multicolumn{6}{l}{$^a$Listed wavelengths are from~\cite{aba,abb}.}\\
\end{tabular}
\end{center}
\label{10T2}
\end{table}

In linear approximation, (\ref{10E15}) can be re-written in the form
\begin{equation}
\frac{\lambda_{i,z}/\lambda_{j,z}}{\lambda_{i,0}/\lambda_{j,0}} =
1 + ({\cal K}_i - {\cal K}_j)\,\Delta\mu/\mu
\label{10E16}
\end{equation}
or
\begin{equation}
z_i = z_0 + \kappa_\mu\,({\cal K}_i - \bar{\cal K})\; ,
\label{10E17}
\end{equation}
where $\lambda_{i,z}$ and $\lambda_{j,z}$ are
the line centers measured in a quasar spectrum, and
\begin{equation}
\kappa_\mu = (1 + z_0)\,\frac{\Delta\mu}{\mu}\; ,
\label{10E18}
\end{equation}
with $z_0$ and $\bar{\cal K}$ being
the mean redshift and the mean sensitivity coefficient, respectively.

The linear regression (\ref{10E17}) calculated for the
complete sample of the H$_2$ lines from Table~2 provides
$(\Delta \mu/\mu)_{J=1+2+3} = (5.0\pm3.2)\times10^{-5}$. 
However, if we consider H$_2$ transitions from individual $J$ levels, then
the weighted mean redshifts reveal \emph{a gradual shift in the radial velocity} 
for features arising from progressively higher rotational levels~:
$z(J$=1) = 3.024890(2), $z(J$=2) = 3.024895(3), and
$z(J$=3) = 3.024904(4).
The H$_2$ lines with changing profiles and small velocity shifts as $J$ increases
were also observed in our Galaxy in the direction of $\zeta$ Ori~A. 
At present these H$_2$ lines are interpreted as being formed 
in \emph{different zones} of a postshock gas~\cite{jp97}.

Although the $z(J$=1) and $z(J$=2) values are consistent
within 1~$\sigma$ intervals, the difference between $z(J$=1) and
$z(J$=3) is essential and equals $1.0\pm0.3$ km~s$^{-1}$.
This small change in the
radial velocity with increasing $J$
may mimic a shift in $\Delta \mu/\mu$.
If we exclude from the regression analysis the rotational levels with $J=3$, then
$(\Delta \mu/\mu)_{J=1+2} = (2.1\pm3.6)\times10^{-5}$.
The use of lines arising from the same rotational levels
would be more reasonable to estimate $\Delta\mu/\mu$,
but in our case the sample size is rather small and we have to combine the
$J=1$ and $J=2$ levels to increase statistics.
The linear regression is shown by the solid line in Fig.~5 while two dashed lines
correspond to the 1~$\sigma$ deviations of the slope parameter $\kappa_\mu$.

Thus, the 1~$\sigma$ confidence interval to $\Delta \mu/\mu$ is 
$ -1.5\times10^{-5} < \Delta \mu/\mu < 5.7\times10^{-5}$.
For a cosmological model with $\Omega_{\rm M} = 0.3$,
$\Omega_\Lambda = 0.7$,
and $H_0 = 72$ km~s$^{-1}$~Mpc$^{-1}$, the look-back time
for $z$ = 3.025 is 11.2 Gyr~\cite{cpt92}.
This leads to the restriction
$|\dot{\mu}/\mu| < 5\times10^{-15}$~yr$^{-1}$
on the variation rate of $\mu$.

\begin{figure}[hpt]
\begin{center}
\includegraphics[width=.99\textwidth]{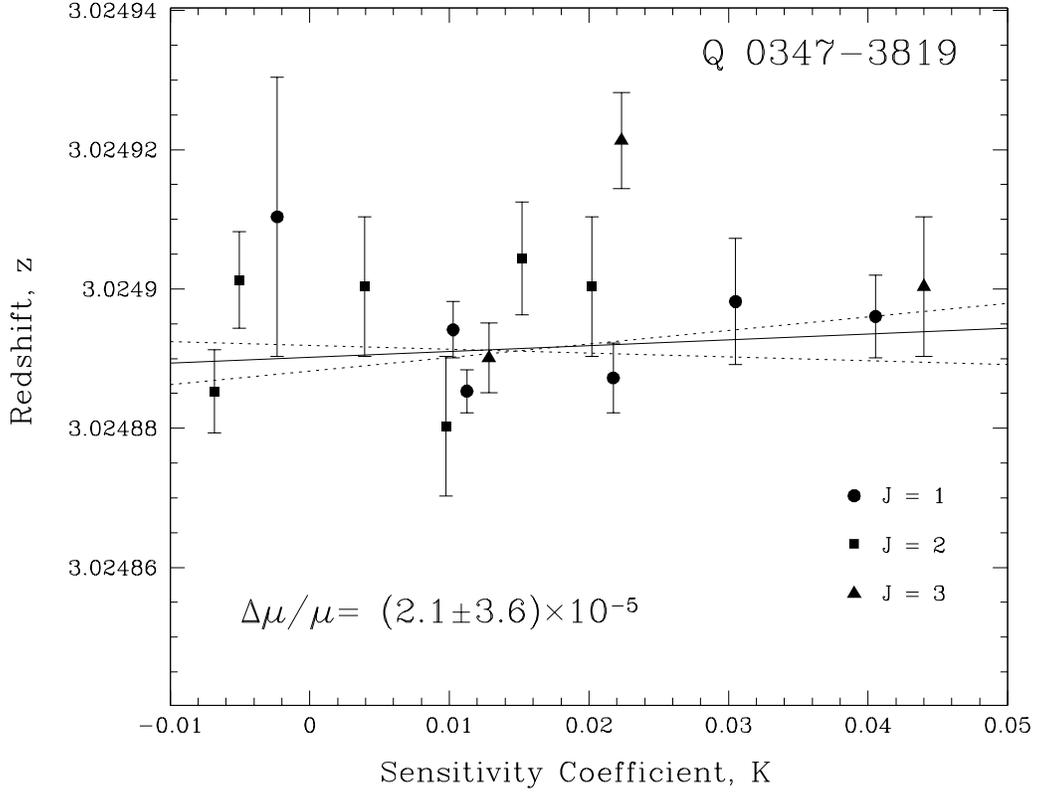}
\end{center}
\caption[]{Relation between the redshift
values $z_i$ calculated for individual H$_2$ lines shown
in Fig.~4 and their sensitivity coefficients ${\cal K}_i$.
The solid line is the linear regression
$z_i = z_0 + \kappa_\mu ({\cal K}_i - \bar{\cal K})$.
The dotted lines representing the 1~$\sigma$ deviations
from the optimal slope $\kappa_\mu$
demonstrate that $\Delta \mu/\mu = 0$ at the level $\sim 5\times10^{-5}$.
}
\label{10F5}
\end{figure}

\section{Conclusions and future prospects}

We have considered modern experimental
investigations of the hypothetical variations of the fundamental
physical constants 
from absorption-line spectra of distant quasars.
The subject was limited to the properties of the two dimensionless
quantities -- the fine-structure constant $\alpha$ and the
proton-to-electron mass ratio $\mu$. 
Our main conclusions are as follows: 
\begin{itemize}
\item
The error $\sigma_{\langle \Delta\alpha/\alpha \rangle} = 1.2\times10^{-6}$
claimed in~\cite{m03} is an order of magnitude smaller compared to what can be
provided by
the wavelength measurements of isolated and unsaturated QSO
absorption lines.
\item
A modification of the MM method (regression MM method)
which makes it more reliable is presented.
This modified MM method applied to the 
homogeneous sample of the isolated and unsaturated Fe\,{\sc ii}
lines yields $\Delta\alpha/\alpha = (1.1\pm1.1)\times10^{-5}$. The 
uncertainty of this value is in agreement with the available
accuracy of the  wavelength measurements. 
\item
If $\Delta\mu/\mu = {\cal R}\,\Delta\alpha/\alpha$, then
the requirement of consistency between $\Delta\alpha/\alpha =
(1.1\pm1.1)\times10^{-5}$ at $z = 1.149$ and $\Delta\mu/\mu =
(2.1\pm3.6)\times10^{-5}$ at $z = 3.025$ rules out very large values
of $|{\cal R}| > 6$, which is in line with the results obtained
in~\cite{No6,No8}.
\end{itemize}

As stated above,
the accuracy of $\sigma_\lambda \sim 0.2$ m\AA\, is required to
achieve a level higher than $10^{-5}$ in the $\Delta\alpha/\alpha$
measurements.  
This accuracy is $\sim 10$ times higher than that available at modern giant
telescopes.
In 2003, a new
High Accuracy Radial velocity Planet Searcher (HARPS) spectrograph~\cite{pepe} 
started to operate at the 3.6~m telescope (ESO, La Silla, Chile).
Observations with HARPS
may in principle provide the desirable accuracy $\sim 0.2$~m\AA.
With a resolving power of 
$\lambda/\Delta\lambda = $ 120\,000 and a projection of the fiber over 4 pixels,
one HARPS pixel corresponds to $\simeq 600$ m~s$^{-1}$. 
For an isolated and unsaturated
absorption line, the line center may be measured with
an accuracy of about $1/20$ of the pixel size
($\sim 30$ m~s$^{-1}$). 
This corresponds to
$\sigma_\lambda = 0.4$ m\AA\, at 4000 \AA.
If isotopic and hyperfine structure systematics as well as the others discussed
in~\cite{m03} are known, then we may hope to reach a level of $\sim 10^{-6}$ in
future astronomical measurements of $\Delta\alpha/\alpha$.

\bigskip\noindent
{\bf Acknowledgments}

\noindent
I thank the organizing committee for providing me with the
opportunity to attend. I especially thank 
Sandro D'Odorico, Luca Pasquini,
Dieter Reimers, and Robert Baade for helpful discussions, and
Xavier Calmet, Keith Olive, and Thomas Dent for
comments on the manuscript.
This work is supported in part by
the RFBR grant No. 03-02-17522.

\end{document}